\documentclass[prl,twocolumn,aps,showpacs]{revtex4-1}

\usepackage{epsfig}
\usepackage{graphicx}
\usepackage{color}
\usepackage{amsmath}

\begin{document}

\title{Equation for the superfluid gap obtained by coarse graining \\ the Bogoliubov-de Gennes equations throughout the BCS-BEC crossover}

\author{S. Simonucci and G. C. Strinati}

\affiliation{Division of Physics, School of Science and Technology \\ Universit\`{a} di Camerino, 62032 Camerino (MC), Italy \\
                 and \\ INFN, Sezione di Perugia, 06123 Perugia (PG), Italy}

\begin{abstract}
We derive a nonlinear differential equation for the gap parameter of a superfluid Fermi system by performing a suitable coarse graining of the Bogoliubov-de Gennes (BdG) equations  throughout the BCS-BEC crossover, with the aim of replacing the time-consuming solution of the original BdG equations by the simpler solution of this novel equation.
We perform a favorable numerical test on the validity of this new equation over most of the temperature-coupling phase diagram, 
by an explicit comparison with the full solution of the original BdG equations for an isolated vortex. 
We also show that the new equation reduces both to the Ginzburg-Landau equation for Cooper pairs in weak coupling close to the critical temperature and to the Gross-Pitaevskii equation for composite bosons in strong coupling at low temperature.
\end{abstract}

\maketitle

%%%%%%%%%%%%%%  INTRODUCTION  %%%%%%%%%%%%%%
% Section 1
\section{I. Introduction} 
The Bogoliubov-de Gennes (BdG) equations \cite{BdG} form the basis for a description of a nouniform Fermi superfluid, and were originally introduced as an extension of the BCS approach \cite{Schrieffer-1964}.
In practice, their numerical solution poses severe problems related to computational time and memory space, since the Pauli principle requires one to obtain a detailed knowledge of a whole set of one-particle eigenfunctions in order to produce eventually the function $\Delta(\mathbf{r})$ representing the spatial dependence of the superfluid gap parameter of interest.
In contrast, superfluidity for bosons (at low temperatures) can be conveniently described by \emph{a single} condensate wave function, which can be directly obtained by solving the Gross-Pitaevskii (GP) differential equation \cite{Gross-1961,Pitaevskii-1961}.

Two cases are already known for which the solution of the BdG equations in nouniform situations can be replaced by the simpler solution of a single differential equation for $\Delta(\mathbf{r})$.
It was shown long ago by Gor'kov \cite{Gorkov-1959} that the Ginzburg-Landau (GL) equation for (largely overlapping) Cooper pairs can be derived from the BdG equations in weak coupling and close to the critical temperature $T_{c}$ at which superfluidity is lost.
More recently, it was shown that the GP equation for composite bosons that form in strong coupling can as well be derived from the BdG equations at low enough temperature \cite{PS-2003}.

In both cases, the microscopic derivations rely on the presence of a small parameter, namely, the ratio $|\Delta(\mathbf{r})|/k_{B}T_{c}$ for the GL equation and the ratio $|\Delta(\mathbf{r})/\mu|$ for the GP equation, where $\mu$ is the chemical potential and $k_{B}$ the Boltzmann constant.
These restrictions limit, in practice, the validity of these differential equations for $\Delta(\mathbf{r})$ to rather small portions of the temperature vs coupling phase diagram.
In the following, we shall use $(k_{F}a_{F})^{-1}$ as the coupling parameter (where $a_{F}$ is the scattering length for two fermions with opposite spins in vacuum and $k_{F}$ is the Fermi wave vector related to the (average) density via $n = k_{F}^{3}/(3 \pi^{2}$)), which ranges from being $\ll -1$ in the weak-coupling (BCS) limit to being $\gg +1$ in the strong-coupling (BEC) limit across the unitary limit where $(k_{F}a_{F})^{-1} = 0$. 

Further attempts have also been made to derive from the BdG equations extensions of the GL equation, which would apply to the BCS regime but at temperatures $T$ somewhat deeper in the superfluid phase away from $T_{c}$ \cite{Tewordt-1963,Werthamer-1963,Werthamer-1964}.
More recently, a systematic expansion of the BdG equations in terms of the small parameter $(T_{c} -T)/T_{c}$ was considered again in the BCS regime \cite{Shanenko-2011}, although it was explicitly tested for the spatially uniform case only.

In this paper, we adopt an alternative strategy and obtain a nonlinear differential equation for the gap parameter $\Delta(\mathbf{r})$
by performing a suitable \emph{coarse graining} of the BdG equations over the microscopic fluctuations of their one-particle eigenfunctions.
Since the smoothness of the spatial variations of the local magnitude \emph{and} phase of the gap parameter $\Delta(\mathbf{r})$ will be the criterion underlying the derivation
of this new equation, we may identify it as a Local Phase Density Approximation (LPDA) to the BdG equations. 
The aim is to replace the solution of the BdG equations themselves by the solution of this simpler equation for $\Delta(\mathbf{r})$ over most of the temperature-coupling phase diagram.
To this end, we will explicitly test the validity of this new equation against the solution of the original BdG equations for the nontrivial case of an isolated vortex, for which a favorable comparison will result over a wide portion of the phase diagram in spite of a considerable reduction of the computation time (that is, a few seconds against a whole day).
This opens the way to possible future applications of the LPDA equation to more complex inhomogeneous situations, for which implementing the BdG equations is essentially out of reach because it is computationally too demanding. 

The paper is organized as follows.
Section II presents a derivation of the LPDA equation for the gap parameter by coarse graining the BdG equations, and shows how both the GL and GP equations can be recovered from the LPDA equation in the appropriate limits. 
A numerical comparison is also presented between the results of the LPDA equation and of the original BdG equations for the case of a single vortex embedded in an infinite superfluid, for several couplings and temperatures.
Section III provides the expressions of the coarse grained number density and current that are consistent with the LPDA approach, and shows a numerical comparison 
with the corresponding BdG results for a single vortex.
Section IV gives our conclusions together with an outlook on possible future applications of the LPDA equation.
In the Appendix analytic expressions are given for the coefficients of the LPDA equation, which are valid at zero temperature throughout the BCS-BEC crossover.

\vspace{-0.2cm}
%%%%%%%%%%%%%%  SECTION 2  %%%%%%%%%%%%%%  
% Section 2
\section{II. The LPDA equation}

In this Section, the LPDA equation for the gap parameter is derived from the original BdG equations, whereby a double coarse graining procedure is introduced for the phase and magnitude of the gap parameter. 
It is also shown that the LPDA equation encompasses the GL equation for largely overlapping Cooper pairs and the GP equation for a dilute gas of composite bosons, which are recovered in the appropriate regions of the temperature-coupling phase diagram.
Numerical results are also presented to test the usefulness and validity of the LPDA equation in practice for a nontrivial case.

%%%%%%%%%%%%%%  Sub-section 2A  %%%%%%%%%%%%%%%%%%%%%%%% 
\vspace{0.1cm}
\begin{center}
{\bf A. Coarse graining the BdG equations}
\end{center}
\vspace{0.1cm}

Formally, the solution of the BdG equations can be written in terms of the associated normal ($\mathcal{G}_{11}$) and anomalous ($\mathcal{G}_{12}$) single-particle GreenÕs functions in the broken-symmetry phase \cite{PS-2003}. 
In particular, the BdG self-consistent equation for the gap parameter takes the form: 
\begin{equation}
- \frac{\Delta(\mathbf{r})^{*}}{v_{0}} = \frac{1}{\beta} \sum_{n} \int \! d\mathbf{r''} \tilde{\mathcal{G}}_{0}(\mathbf{r''},\mathbf{r};-\omega_{n}) 
\Delta(\mathbf{r''})^{*} \mathcal{G}_{11}(\mathbf{r''},\mathbf{r};\omega_{n})
\label{BdG-gap-equation}
\end{equation}
\noindent
where $v_{0}$ is the strength of the attractive interparticle interaction of the contact type, $\omega_{n} = k_{B} T (2n+1)\pi$ ($n$ integer) is a Matsubara frequency, and $\tilde{\mathcal{G}}_{0}$ is the noninteracting Green's function that satisfies the equation:
\begin{equation}
\left[ i \omega_{n} \, - \, \mathcal{H}(\mathbf{r}) \right] \, 
\tilde{\mathcal{G}}_{0}(\mathbf{r},\mathbf{r'};\omega_{n}) \, = \, \delta(\mathbf{r} - \mathbf{r'}) \, .
\label{noninteracting-equation}
\end{equation}
\noindent
Here, $ \mathcal{H}(\mathbf{r}) = (i \nabla + \mathbf{A}(\mathbf{r}))^{2}/(2m) + V(\mathbf{r}) - \mu$ contains the vector potential $\mathbf{A}(\mathbf{r})$ (in the Coulomb gauge) as well as an external potential $V(\mathbf{r})$ (we set $\hbar =1$ and \emph{e} = 1).
Accordingly, in what follows it is convenient to introduce a local chemical potential $\bar{\mu}(\mathbf{r}) = \mu - V(\mathbf{r}) - \mathbf{A}(\mathbf{r})^{2}/(2m)$. 
[For neutral atoms in a rotating trap, for which $\mathbf{A}(\mathbf{r}) = m \, \mathbf{\Omega}  \wedge \mathbf{r}$ where $\mathbf{\Omega}$ is the angular velocity, $\bar{\mu}(\mathbf{r})$
does not contain the term $\propto \mathbf{A}^{2}$.]

% Figure 1
\begin{figure}[h]
\includegraphics[angle=0,width=8.7cm]{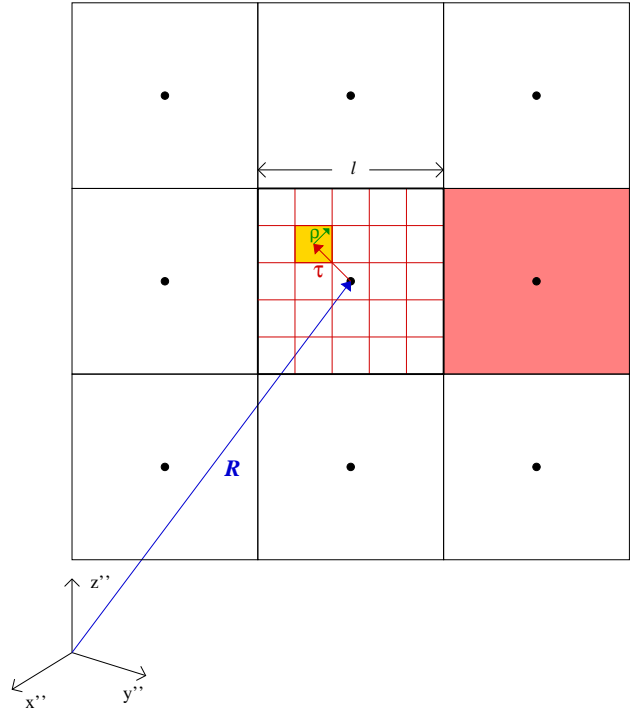}
\caption{(Color online) Double coarse graining procedure. Volumes of side $\ell$ centered at $\mathbf{R}$ are identified where the magnitude $\tilde{\Delta}(\mathbf{R})$ 
of the gap parameter is considered to be (approximately) constant. Embedded in them, smaller volumes centered at $\mathbf{R} + \boldsymbol{\tau}$ are further identified where (the gradient of) the phase $2 \mathbf{Q}(\mathbf{R},\boldsymbol{\tau})$ of the gap parameter is also considered to be (approximately) constant.}
\label{Figure-1}
\end{figure}

The coarse graining of the gap equation (\ref{BdG-gap-equation}) proceeds as follows.
The variable $\mathbf{r''}$ in Eq.(\ref{BdG-gap-equation}) is written as $\mathbf{r''} = \mathbf{R} + \boldsymbol{\tau} + \boldsymbol{\rho}$, where $\mathbf{R}$ and 
$\boldsymbol{\tau}$ identify, in order, the centers of the volume elements (embedded into one another) about which the magnitude 
$\tilde{\Delta}(\mathbf{R})$ and (the gradient of) the phase $2 \mathbf{Q}(\mathbf{R},\boldsymbol{\tau})$ of the gap are considered to be approximately constant (cf. Fig.÷\ref{Figure-1}). 
We write:
\begin{equation}
\Delta(\mathbf{r''}) = \tilde{\Delta}(\mathbf{R}) \, e^{2 i \mathbf{Q}(\mathbf{R},\boldsymbol{\tau}) \cdot (\mathbf{R} + \boldsymbol{\tau} + \boldsymbol{\rho})} \, .
\label{magnitude_and_phase}
\end{equation}
\noindent
Locally in the smaller volume element centered at $\mathbf{R}+\boldsymbol{\tau}$, the problem is then equivalent to a Fulde-Ferrell phase \cite{FF-1964} with balanced spin populations and wave vector $\mathbf{Q}(\mathbf{R},\boldsymbol{\tau})$, so that in Eq.(\ref{BdG-gap-equation}):
\begin{equation}
\mathcal{G}_{11}(\mathbf{r''},\mathbf{r};\omega_{n}) = 
e^{ i \mathbf{Q}(\mathbf{R},\boldsymbol{\tau}) \cdot (\mathbf{R} + \boldsymbol{\tau} + \boldsymbol{\rho} - \mathbf{r})} \,
\mathcal{G}_{11}^{\mathbf{A}} (\mathbf{R} + \boldsymbol{\tau} + \boldsymbol{\rho} - \mathbf{r};\omega_{n}|\mathbf{r})
\label{FF-Green-function-I}
\end{equation}
\noindent
where we have assumed that the volume element centered at $\mathbf{R}+\boldsymbol{\tau}$ is close to the variable $\mathbf{r}$ in Eq.(\ref{BdG-gap-equation}), 
and
\begin{equation}
\mathcal{G}_{11}^{\mathbf{A}} (\mathbf{x};\omega_{n}) \! = \! \!
\int \! \frac{d\mathbf{k}}{(2 \pi)^{3}} 
\frac{e^{i \mathbf{k} \cdot \mathbf{x}} \, [i \omega_{n} + \xi^{\mathbf{A}}(\mathbf{k}-\mathbf{Q})]}
{(i \omega_{n} - E_{+}^{\mathbf{A}}(\mathbf{k};\mathbf{Q})) (i \omega_{n} + E_{-}^{\mathbf{A}}(\mathbf{k};\mathbf{Q}))} \, ,
\label{FF-Green-function-II}
\end{equation}
\noindent
with $\xi^{\mathbf{A}}(\mathbf{k}-\mathbf{Q}) = \frac{(\mathbf{k}-\mathbf{Q})^{2}}{(2m)} - \bar{\mu} + \mathbf{A}\cdot \frac{(\mathbf{k}-\mathbf{Q})}{m}$ and
\begin{eqnarray}
E_{\pm}^{\mathbf{A}}(\mathbf{k};\mathbf{Q}) & = &
\sqrt{\left( \frac{\mathbf{k}^{2}}{2m} + \frac{\mathbf{Q}^{2}}{2m} - \bar{\mu} - \frac{\mathbf{A}}{m} \cdot \mathbf{Q} \right)^{2} + \tilde{\Delta}^{2}} 
\nonumber \\
& \pm & \frac{\mathbf{k}}{m} \cdot (\mathbf{Q}-\mathbf{A}) \, .
\label{definition-E-pm}
\end{eqnarray}
\noindent
When the expression (\ref{FF-Green-function-II}) is used in Eq.(\ref{FF-Green-function-I}), local values $\bar{\mu}(\mathbf{r})$ and $\mathbf{A}(\mathbf{r})$ are there implied (as indicated by the notation $|\mathbf{r})$ in Eq.(\ref{FF-Green-function-I})).
Similarly, we write for the noninteracting counterpart:
\begin{equation}
\tilde{\mathcal{G}}_{0}(\mathbf{r''},\mathbf{r};-\omega_{n}) = \int \! \frac{d\mathbf{k}}{(2 \pi)^{3}} 
\frac{e^{i \mathbf{k} \cdot (\mathbf{r''} - \mathbf{r})}}
{-i\omega_{n} - \frac{\mathbf{k}^{2}}{2m} +  \frac{\mathbf{A(\mathbf{r})}}{m} \cdot \mathbf{k} + \bar{\mu}(\mathbf{r})}
\label{noninteracting-counterpart}
\end{equation}
\noindent
in terms of a (local) eikonal approximation \cite{Gorkov-1959}. 

In this way, upon integrating over $\boldsymbol{\rho}$ and summing over $\omega_{n}$, from Eq.(\ref{BdG-gap-equation}) one arrives at the expression:
\begin{eqnarray}
& & - \frac{\Delta(\mathbf{r})^{*}}{v_{0}} = \sum_{ \{ \mathbf{R} \} } \, \tilde{\Delta}(\mathbf{R}) \, \sum_{ \{ \boldsymbol{\tau} \} } \,
e^{- 2 i \mathbf{Q}(\mathbf{R},\boldsymbol{\tau}) \cdot \mathbf{r}}  
\nonumber \\
& \times & \int \! \frac{d\mathbf{k}}{(2 \pi)^{3}}
\frac{ 1 - 2 \, f_{F}(E_{+}^{\mathbf{A}}(\mathbf{k};\mathbf{Q}(\mathbf{R},\boldsymbol{\tau})|\mathbf{r})) }
{2 E^{\mathbf{A}}(\mathbf{k};\mathbf{Q}(\mathbf{R},\boldsymbol{\tau})|\mathbf{r})}
\label{intermediate-LPDA-equation}
\end{eqnarray}
\noindent
where $f_{F}(E) = \left( e^{E/(k_{B}T)} + 1 \right)^{-1}$ is the Fermi function and $2 E^{\mathbf{A}}(\mathbf{k};\mathbf{Q}|\mathbf{r}) = E_{+}^{\mathbf{A}}(\mathbf{k};\mathbf{Q}|\mathbf{r}) + E_{-}^{\mathbf{A}}(\mathbf{k};\mathbf{Q}|\mathbf{r})$.

At this point, further approximations involve: 
(i) Setting in the exponent 
$\mathbf{Q}(\mathbf{R},\boldsymbol{\tau}) \cdot \mathbf{r} \simeq \mathbf{Q}(\mathbf{R},\boldsymbol{\tau}=0) \cdot \mathbf{R} + 
\mathbf{Q}(\mathbf{R},\boldsymbol{\tau}) \cdot (\mathbf{r} - \mathbf{R})$;
(ii) Transforming the sum over $\boldsymbol{\tau}$ into an integral over the independent variable $\mathbf{Q}$ under the assumption that all the relevant values of $\mathbf{Q}$ are effectively sampled by varying $\boldsymbol{\tau}$ in the volume of side $\ell$ centered at $\mathbf{R}$ (cf. Fig.÷\ref{Figure-1});
(iii) Transforming also the sum over $\mathbf{R}$ into an integral;
(iv) Identifying $\Delta(\mathbf{R}) = e^{2 i \mathbf{Q}(\mathbf{R},\boldsymbol{\tau}=0) \cdot \mathbf{R}} \tilde{\Delta}(\mathbf{R})$;
(v) Eliminating $v_{0}$ in favor of $a_{F}$ through a standard regularization \cite{Randeria-1993,PS-2000}.
The gap equation then becomes:
\begin{equation}
- \frac{m}{4 \pi a_{F}} \, \Delta(\mathbf{r}) =  \! \int \! d \mathbf{R} \,\, \Delta(\mathbf{R}) \! \int \! \! \frac{d\mathbf{Q}}{\pi^{3}} \, 
e^{2 i \mathbf{Q} \cdot (\mathbf{r} - \mathbf{R})} \, K^{\mathbf{A}}(\mathbf{Q}|\mathbf{r})
\label{not-yet-final-LPDA-equation}
\end{equation}
\noindent
where we have introduced the kernel \cite{footnote-kernel}

\begin{equation}
K^{\mathbf{A}}(\mathbf{Q}|\mathbf{r}) \! = \! \! \int \! \frac{d\mathbf{k}}{(2 \pi)^{3}} 
\left\{ \frac{ 1 - 2 \, f_{F}(E_{+}^{\mathbf{A}}(\mathbf{k};\mathbf{Q}|\mathbf{r}))}
{2 E^{\mathbf{A}}(\mathbf{k};\mathbf{Q}|\mathbf{r})} - \frac{m}{\mathbf{k}^{2}} \right\} \, .
\label{definition-relevant-kernel} 
\end{equation}

The desired differential equation for $\Delta(\mathbf{r})$ results eventually from Eq.(\ref{not-yet-final-LPDA-equation}) by expanding the kernel $ K^{\mathbf{A}}(\mathbf{Q}|\mathbf{r})$ in powers of $\mathbf{Q}$ and integrating by parts the integral over $\mathbf{R}$ therein.
Up to quadratic order, one obtains \cite{footnote-negligible}:
\begin{eqnarray}
- \frac{m}{4 \pi a_{F}} \, \Delta(\mathbf{r}) & = &  \mathcal{I}_{0}(\mathbf{r}) \, \Delta(\mathbf{r}) + 
\mathcal{I}_{1}(\mathbf{r}) \, \frac{\nabla^{2}}{4m} \Delta(\mathbf{r})                    
\nonumber \\ 
& - & \mathcal{I}_{1}(\mathbf{r}) \, i \, \frac{\mathbf{A}(\mathbf{r})}{m} \cdot \nabla \Delta(\mathbf{r})                                                          
\label{LPDA-differential-equation}
\end{eqnarray}
\noindent
with the notation
\begin{equation}
\mathcal{I}_{0}(\mathbf{r}) = \int \! \frac{d \mathbf{k}}{(2 \pi)^{3}} \, 
\left\{ \frac{ 1 - 2 f_{F}(E_{+}^{\mathbf{A}}(\mathbf{k}|\mathbf{r})) }{2 \, E(\mathbf{k}|\mathbf{r})} - \frac{m}{\mathbf{k}^{2}} \right\}
\label{I_0-definition-finite_temperature}
\end{equation}
\noindent
and
\begin{eqnarray}
\mathcal{I}_{1}(\mathbf{r}) & = & \frac{1}{2} \, \int \! \frac{d \mathbf{k}}{(2 \pi)^{3}} 
\left\{  \frac{\xi(\mathbf{k}|\mathbf{r})}{2 \, E(\mathbf{k}|\mathbf{r})^{3}} \, \left[  1 - 2 f_{F}(E_{+}^{\mathbf{A}}(\mathbf{k}|\mathbf{r})) \right] \right.                          
\nonumber \\
& + &  \frac{\xi(\mathbf{k}|\mathbf{r})}{E(\mathbf{k}|\mathbf{r})^{2}} \, 
\frac{\partial f_{F}(E_{+}^{\mathbf{A}}(\mathbf{k}|\mathbf{r}))}{\partial E_{+}^{\mathbf{A}}(\mathbf{k}|\mathbf{r})} 
\label{I_1-definition-finite_temperature} \\
& - & \left. \frac{\mathbf{k}\cdot\mathbf{A}(\mathbf{r})}{\mathbf{A}(\mathbf{r})^{2}} \, \frac{1}{E(\mathbf{k}|\mathbf{r})} \, 
\frac{\partial f_{F}(E_{+}^{\mathbf{A}}(\mathbf{k}|\mathbf{r}))}{\partial E_{+}^{\mathbf{A}}(\mathbf{k}|\mathbf{r})} \right\} 
\nonumber
\end{eqnarray}
\noindent
where 
$\xi(\mathbf{k}|\mathbf{r}) = \frac{\mathbf{k}^{2}}{2m} - \bar{\mu}(\mathbf{r})$,
$E(\mathbf{k}|\mathbf{r}) = \sqrt{\xi(\mathbf{k}|\mathbf{r})^{2} + |\Delta(\mathbf{r})|^{2}}$, and
$E_{+}^{\mathbf{A}}(\mathbf{k}|\mathbf{r}) = E(\mathbf{k}|\mathbf{r}) - \frac{\mathbf{k} \cdot \mathbf{A}(\mathbf{r})}{m}$. 

Equation (\ref{LPDA-differential-equation}) represents the main result of the present paper.
From the way it was obtained, we may regard it as representing a Local Phase Density Approximation (LPDA), that should hold with \emph{no a priori restrictions on coupling and temperature regimes}, provided that $\Delta(\mathbf{r})$ varies slowly enough with its magnitude varying more slowly than its phase. 
Note, in particular, the presence of the vector potential in the arguments of the Fermi functions entering the coefficients (\ref{I_0-definition-finite_temperature}) and 
(\ref{I_1-definition-finite_temperature}) of the LPDA equation, which results in a kind of a local Fulde-Ferrell phase.
This feature (which will be a crucial ingredient when applying the LPDA equation, for instance, to neutral fermions in a rotating trap) distinguishes, too, the present from other proposals also based on the slow spatial variation of the gap parameter \cite{Devreese-2013}.

In addition, for a sufficiently small $\mathbf{A}(\mathbf{r})$ one may expand $f_{F}(E_{+}^{\mathbf{A}}(\mathbf{k}|\mathbf{r}))$ and 
$\partial f_{F}(E_{+}^{\mathbf{A}}(\mathbf{k}|\mathbf{r})) / \partial E_{+}^{\mathbf{A}}(\mathbf{k}|\mathbf{r})$ in Eqs.(\ref{I_0-definition-finite_temperature}) and (\ref{I_1-definition-finite_temperature}) in powers of $\mathbf{k} \cdot \mathbf{A}(\mathbf{r})$.
In this case:
\begin{eqnarray}
& & \mathcal{I}_{1}(\mathbf{r}) \cong \frac{1}{2} \, \int \! \frac{d \mathbf{k}}{(2 \pi)^{3}} 
\left\{  \frac{\xi(\mathbf{k}|\mathbf{r})}{2 \, E(\mathbf{k}|\mathbf{r})^{3}} \, \left[  1 - 2 f_{F}(E(\mathbf{k}|\mathbf{r})) \right] \right.                          
\nonumber \\
& + &  \left. \frac{\xi(\mathbf{k}|\mathbf{r})}{E(\mathbf{k}|\mathbf{r})^{2}} \, 
\frac{\partial f_{F}(E(\mathbf{k}|\mathbf{r}))}{\partial E(\mathbf{k}|\mathbf{r})} +
\frac{\mathbf{k}^{2}/(3m)}{E(\mathbf{k}|\mathbf{r})} \, 
\frac{\partial^{2} f_{F}(E(\mathbf{k}|\mathbf{r}))}{\partial E(\mathbf{k}|\mathbf{r})^{2}} \right\}  
\nonumber
\end{eqnarray}
\noindent
and
\[
\mathcal{I}_{0}(\mathbf{r})  \! \cong  \! \int \! \frac{d \mathbf{k}}{(2 \pi)^{3}}  \! 
\left\{ \frac{\left[  1 - 2 f_{F}(E(\mathbf{k}|\mathbf{r})) \right]}{2 \, E(\mathbf{k}|\mathbf{r})} - \frac{m}{\mathbf{k}^{2}} \right\} - \frac{\mathbf{A}(\mathbf{r})^{2}}{m} \mathcal{I}_{1}(\mathbf{r}) 
\]
\noindent
where now the local chemical potential $\mu(\mathbf{r}) = \mu - V(\mathbf{r})$ no longer contains the $\mathbf{A}(\mathbf{r})^{2}$ term.
Grouping all terms containing $\mathcal{I}_{1}(\mathbf{r})$ in Eq.(\ref{LPDA-differential-equation}), one correctly recovers the gauge-invariant form 
$-\frac{\mathcal{I}_{1}(\mathbf{r})}{4m} (i\nabla+2\mathbf{A}(\mathbf{r}))^{2}$.
Related expressions for the coarse grained number density and current will be obtained in Section III.

%%%%%%%%%%%%%%  Sub-section 2B  %%%%%%%%%%%%%%%%%%%%%%%% 
\vspace{0.1cm}
\begin{center}
{\bf B. Recovering the GL and GP equations}
\end{center}
\vspace{0.1cm}

The LPDA equation reduces to the GL and GP equations in the appropriate limits, which can be shown as follows.

For weak coupling $(k_{F}a_{F})^{-1} \ll -1$ and temperatures close to $T_{c}$, in the above expression for 
$\mathcal{I}_{1}(\mathbf{r})$ one can approximate $E(\mathbf{k}|\mathbf{r}) \cong |\xi(\mathbf{k}|\mathbf{r})|$ and neglect the terms whose integrands are odd in 
$\xi(\mathbf{k}|\mathbf{r})$.
Omitting further the external potential, one obtains 
$\mathcal{I}_{1}(\mathbf{r}) \cong \frac{k_{F}^{2}}{2m} \frac{N_{0}}{6 (k_{B}T_{c})^{2}} \int_{0}^{\infty} \! \frac{d y}{y} \frac{\tanh y}{\cosh^{2} y}$ where $N_{0} = m k_{F}/(2 \pi^{2})$ is the density of states at the Fermi level per spin component.
In addition, using the BCS equation for $T_{c}$ one obtains $\mathcal{I}_{0}(\mathbf{r}) \cong - \frac{m}{4 \pi a_{F}} + N_{0} \frac{(T_{c} - T)}{T_{c}} -
 \frac{7 \, \zeta(3)}{8 \pi^{2}} \frac{N_{0}}{(k_{B}T_{c})^{2}} |\Delta(\mathbf{r})|^{2}$ where $\zeta(3)$ is the Riemann zeta function of argument $3$.
 The GL equation is thus readily recovered from the LPDA equation (\ref{LPDA-differential-equation}) in this limit \cite{Gorkov-1959}.

In the opposite limit of strong coupling $(k_{F}a_{F})^{-1} \gg +1$ and low temperatures, the two-body binding energy 
$\varepsilon_{0} = (m a_{F}^{2})^{-1} = - 2 \mu + \mu_{B}$ is the largest energy scale in the problem, where $\mu_{B}$ is the residual chemical potential for the composite bosons that form in this limit.
To the leading significant order, one obtains $\mathcal{I}_{1}(\mathbf{r}) \cong \frac{m^{2} a_{F}}{8 \pi}$ and
$\mathcal{I}_{0}(\mathbf{r}) \cong - \frac{m}{4 \pi a_{F}} + \frac{m^{2} a_{F}}{8 \pi} [ \mu_{B} - 2 \, V(\mathbf{r}) -  \frac{m a_{F}^{2}}{2 \pi} |\Delta(\mathbf{r})|^{2}]$.
The GP equation for composite bosons is thus readily recovered  from the LPDA equation (\ref{LPDA-differential-equation}) in this limit \cite{PS-2003}.

%%%%%%%%%%%%%%  Sub-section 2C  %%%%%%%%%%%%%%%%%%%%%%%% 
\vspace{0.1cm}
\begin{center}
{\bf C. Numerical comparison for an isolated vortex}
\end{center}
\vspace{0.1cm}

We pass now to test the numerical solution of the LPDA equation (\ref{LPDA-differential-equation}) with $\mathbf{A}(\mathbf{r}) = 0$ for the nontrivial case of an isolated vortex embedded in an infinite medium, against the results of the accurate solution of the BdG equations reported in Ref.\cite{SPS-2013} across the BCS-BEC crossover for all $ T < T_{c}$.
This case exemplifies the situation depicted in Fig.÷\ref{Figure-1}, whereby the magnitude of the gap parameter varies more slowly than its phase, and actually represents a rather extreme situation since the gradient of the phase diverges when approaching the center of the vortex.

\begin{figure}[t]
\includegraphics[angle=0,width=8.7cm]{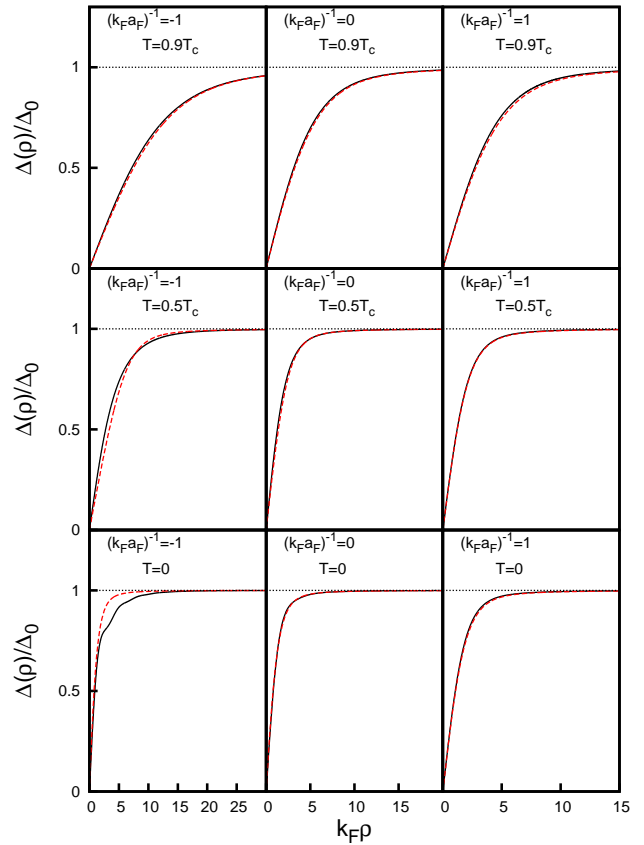}
\caption{(Color online) Radial profiles of the gap parameter $\Delta(\rho)$ for an isolated vortex embedded in an infinite fermionic superfluid, for various temperatures and couplings. In each case, the results obtained by solving the LPDA equation (\ref{LPDA-differential-equation}) (dashed lines) are compared with those obtained by the full solution of the BdG equations obtained in Ref.\cite{SPS-2013} (full lines).}
\label{Figure-2}
\end{figure}

Figure \ref{Figure-2} shows the profiles $\Delta(\rho)$ of the gap parameter (in units of the asymptotic value $\Delta_{0}$ away from the center of the vortex)
vs the radial distance $\rho$ (in units of $k_{F}^{-1}$) for various temperatures and couplings across unitarity,  obtained by solving the LPDA equation (\ref{LPDA-differential-equation}) (dashed lines) and from the BdG calculation of Ref.\cite{SPS-2013} (full lines).

% Figure 2

In all these cases, the overall agreement between the two calculations appears to be extremely good, considering also the fact that the coherence (healing) length changes substantially from case to case, and appears especially remarkable in the light of the huge reduction of computational time (by a factor of about $10^{5}$) that results in the LPDA calculation with respect to the BdG calculation.
Deviations between the two calculations emerge essentially in the BCS regime at low temperature, where the LPDA calculation fails to reproduce the Friedel's oscillations that are present in the BdG calculation over the microscopic length scale $k_{F}^{-1}$ (which has been ``coarse-grained" by the LPDA approach).
That a local differential approach might be bound to fail in the BCS regime at low temperature was already pointed out in Refs.\cite{Tewordt-1963}-\cite{Werthamer-1964}, but was never explicitly verified against a nontrivial benchmark like the BdG calculation here considered. 
The reason for the failure of a local differential approach in the BCS regime at low temperature should be traced in the spatial range of the kernel from which this differential equation is obtained in the final step, since this range (which is of the order of the size of the fermion pairs at low temperature) about coincides with the range of the gap parameter itself, thus limiting the validity of a local (differential) approach.

\vspace{-0.2cm}
%%%%%%%%%%%%%%  SECTION 3  %%%%%%%%%%%%%%  
% Section 3
\section{III. Coarse grained density and current} 

In this Section, we provide additional information about the expressions of the number density and current which are consistent with the LPDA approach developed in Section II.

%%%%%%%%%%%%%%  Sub-section 3A  %%%%%%%%%%%%%%%%%%%%%%%% 
\vspace{0.1cm}
\begin{center}
{\bf A. Coarse grained density}
\end{center}
\vspace{0.1cm}

With reference to Fig.\ref{Figure-1} and Eq.(\ref{FF-Green-function-II}), the number density at a point $\mathbf{r}$ inside the small volume element centered at $\mathbf{R}+\boldsymbol{\tau}$, to which there corresponds the wave vector $\mathbf{Q}(\mathbf{R},\boldsymbol{\tau})$, has the form:
\begin{eqnarray}
n(\mathbf{r}) & = & 2 k_{B} T \sum_{n} e^{i \omega_{n} \eta} \mathcal{G}_{11}(\mathbf{r},\mathbf{r};\omega_{n})
\nonumber \\
& \rightarrow & 2 k_{B} T \sum_{n} e^{i \omega_{n} \eta} \mathcal{G}^{\mathbf{A}}_{11}(0;\omega_{n},\mathbf{Q}(\mathbf{R},\boldsymbol{\tau})|\mathbf{r})
\nonumber \\
& = & \int \! \frac{d\mathbf{k}}{(2 \pi)^{3}} \left\{ 1 - 
\frac{\xi^{\mathbf{A}}(\mathbf{k};\mathbf{Q}(\mathbf{R},\boldsymbol{\tau})|\mathbf{r})}{E^{\mathbf{A}}(\mathbf{k};\mathbf{Q}(\mathbf{R},\boldsymbol{\tau})|\mathbf{r})} \right.
\nonumber \\
& \times & \left. \left[ 1 - 2 f_{F}(E^{\mathbf{A}}_{+}(\mathbf{k};\mathbf{Q}(\mathbf{R},\boldsymbol{\tau})|\mathbf{r})) \right] \right\} 
\label{density-general}
\end{eqnarray}
\noindent
where $\eta$ is a positive infinitesimal and
\begin{eqnarray}
\xi^{\mathbf{A}}(\mathbf{k};\mathbf{Q}(\mathbf{R},\boldsymbol{\tau})|\mathbf{r}) & = & \frac{\mathbf{k}^{2}}{2m} - \mu(\mathbf{r}) + 
\frac{(\mathbf{Q}(\mathbf{R},\boldsymbol{\tau}) - \mathbf{A}(\mathbf{r}))^{2}}{2m}\, ,
\nonumber \\
E^{\mathbf{A}}(\mathbf{k};\mathbf{Q}(\mathbf{R},\boldsymbol{\tau})|\mathbf{r}) & = & \sqrt{\xi^{\mathbf{A}}(\mathbf{k};\mathbf{Q}(\mathbf{R},\boldsymbol{\tau})|\mathbf{r})^{2}
+ |\Delta(\mathbf{r})|^{2}}  \, ,
\nonumber \\
E^{\mathbf{A}}_{+}(\mathbf{k};\mathbf{Q}(\mathbf{R},\boldsymbol{\tau})|\mathbf{r}) & = & E^{\mathbf{A}}(\mathbf{k};\mathbf{Q}(\mathbf{R},\boldsymbol{\tau})|\mathbf{r}) 
\nonumber \\
& + & \frac{\mathbf{k}}{m} \cdot (\mathbf{Q}(\mathbf{R},\boldsymbol{\tau}) - \mathbf{A}(\mathbf{r})) \, .
\label{three-definitions}
\end{eqnarray}
\noindent
In the above expressions, $\mu(\mathbf{r}) = \mu - V(\mathbf{r})$ contains only the external potential $V(\mathbf{r})$ and the $\mathbf{r}$ dependence
originates from the local values of $\mu(\mathbf{r})$ and $\Delta(\mathbf{r})$.
Recalling, in addition, that the wave vector $\mathbf{Q}(\mathbf{R},\boldsymbol{\tau})$ is associated with $\nabla \varphi(\mathbf{r}) / 2$ where $ \varphi(\mathbf{r})$ is the phase of the gap parameter $\Delta(\mathbf{r}) = |\Delta(\mathbf{r})| e^{i \varphi(\mathbf{r})}$, it is useful to rewrite the above LPDA expression for 
$n(\mathbf{r})$ in the more standard form:
\begin{equation}
n(\mathbf{r}) =  \int \! \frac{d\mathbf{k}}{(2 \pi)^{3}} \left\{ 1 - 
\frac{\xi^{\mathbf{A}}(\mathbf{k}|\mathbf{r})}{E^{\mathbf{A}}(\mathbf{k}|\mathbf{r})} 
\left[ 1 - 2 f_{F}(E^{\mathbf{A}}_{+}(\mathbf{k}|\mathbf{r})) \right] \right\} 
\label{number-density-general}
\end{equation}
\noindent
where
\begin{eqnarray}
\xi^{\mathbf{A}}(\mathbf{k}|\mathbf{r}) & = & \frac{\mathbf{k}^{2}}{2m} - \mu(\mathbf{r}) + 
\frac{1}{2m} \left( \frac{\nabla \varphi(\mathbf{r})}{2} - \mathbf{A}(\mathbf{r}) \right)^{2} \, ,
\nonumber \\
E^{\mathbf{A}}(\mathbf{k}|\mathbf{r}) & = & \sqrt{\xi^{\mathbf{A}}(\mathbf{k}|\mathbf{r})^{2}
+ |\Delta(\mathbf{r})|^{2}}  \, ,
\nonumber \\
E^{\mathbf{A}}_{+}(\mathbf{k}|\mathbf{r}) & = & E^{\mathbf{A}}(\mathbf{k}|\mathbf{r}) 
+ \frac{\mathbf{k}}{m} \cdot\left( \frac{\nabla \varphi(\mathbf{r})}{2} - \mathbf{A}(\mathbf{r}) \right) \, .
\label{three-definitions}
\end{eqnarray}

This expression for the local density can even be used in the central region of a vortex where $|\Delta(\mathbf{r})| \rightarrow 0$ but $\nabla \varphi(\mathbf{r}) \rightarrow \infty$ at the same time (to deal with this case, we set $V(\mathbf{r})=0$ and $A(\mathbf{r})=0$).
The way different terms act in Eq.(\ref{number-density-general}) can be most readily understood in the case of zero temperature.
If one neglects the presence of $\nabla \varphi(\mathbf{r})$ altogether in the expression (\ref{number-density-general}), the density reduces to the Local Density Approximation (LDA) form:
\begin{equation}
\bar{n}(\mathbf{r}) = \int \! \frac{d\mathbf{k}}{(2 \pi)^{3}} 
\left\{ 1 - \frac{\xi(\mathbf{k}|\mathbf{r})}{E(\mathbf{k}|\mathbf{r})} \left[ 1 - 2 f_{F}(E(\mathbf{k}|\mathbf{r})) \right] \right\} 
\label{coarse-garined-density}
\end{equation}
\noindent
where 
\vspace{-0.2cm}
\begin{eqnarray}
\xi(\mathbf{k}|\mathbf{r}) & = & \frac{\mathbf{k}^{2}}{2m} - \mu \, ,
\nonumber \\
E(\mathbf{k}|\mathbf{r}) & = & \sqrt{\xi(\mathbf{k}|\mathbf{r})^{2} + |\Delta(\mathbf{r})|^{2}}  \, .
\label{two-definitions}
\end{eqnarray}
\vspace{0.01cm}
\noindent
At the center of the vortex where $|\Delta(\mathbf{r})| = 0$, the value of $\bar{n}(\mathbf{r}=0) = k_{\mu}^{3}/(3 \pi^{2})$ (where $k_{\mu} = \sqrt{2 m \mu}$ when  $\mu > 0$ and zero otherwise) corresponds to the density of a noninteracting Fermi gas with the value $\mu$ for the Fermi energy.
Replacing then the chemical potential $\mu$ by the local value $\mu - \frac{(\nabla \varphi(\mathbf{r})/2)^{2}}{2m}$ like in the expression (\ref{three-definitions}) for
$\xi^{\mathbf{A}}(\mathbf{k}|\mathbf{r})$ brings this value down to zero for (positive) $\mu$.
But as soon as the effect of $\nabla \varphi(\mathbf{r})$ is restored also in the last term of the expression (\ref{three-definitions}) for $E^{\mathbf{A}}_{+}(\mathbf{k}|\mathbf{r})$, the effect of the Fermi function in Eq.(\ref{number-density-general}) is to bring the value of $n(\mathbf{r}=0)$ back to 
$k_{\mu}^{3}/(3 \pi^{2})$.

To prove this statement, we set $\mathbf{Q} = \nabla \varphi(\mathbf{r})/2$ (where $|\mathbf{Q}| \rightarrow \infty$ at the end of the calculation) and consider in Eq.(\ref{number-density-general}) the (positive) contribution of the term that contains the Fermi function with $|\Delta(\mathbf{r})| = 0$.
In this way we obtain: 
\begin{eqnarray}
& & 2 \int \! \frac{d\mathbf{k}}{(2 \pi)^{3}} \, f_{F}(E^{\mathbf{A}}_{+}(\mathbf{k}|\mathbf{r})) 
\label{contribution_from_Fermi-function} \\
& = & 2 \int \! \frac{d\mathbf{k}}{(2 \pi)^{3}} \, f_{F} \!\! \left(\frac{k^{2}}{2m} - \mu + \frac{Q^{2}}{2m} + \frac{\mathbf{k} \cdot \mathbf{Q}}{m} \right)
\nonumber \\
& = & \frac{1}{2 \pi^{2}} \int_{Q-k_{\mu}}^{Q+k_{\mu}} \!\!  dk \, k^{2} \left[ 1 - \frac{\left(\frac{k^{2}}{2m} - \mu + \frac{Q^{2}}{2m}\right)}{\frac{k \, Q}{m}} \right]
= \frac{k_{\mu}^{3}}{3 \pi^{2}} 
\nonumber
\end{eqnarray}
\noindent
as anticipated.
As the temperature is increased above zero, on the other hand, the term in Eq.(\ref{number-density-general}) containing the Fermi function becomes progressively more important even on the BEC side of unitarity when $\mu < 0$.

% Figure 3
\begin{figure}[t]
\includegraphics[angle=0,width=8.7cm]{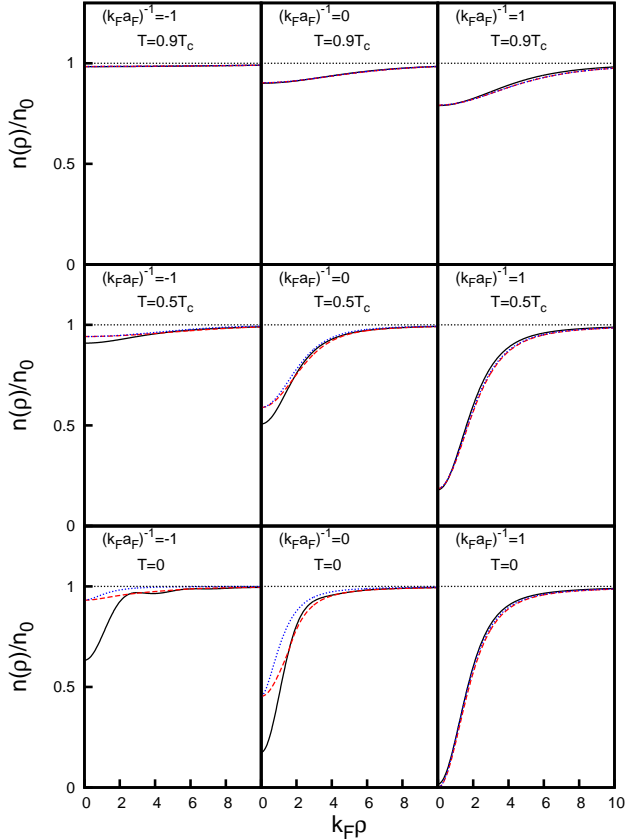}
\caption{(Color online) Radial profiles of the number density (in units of its asymptotic value $n_{0}$) for an isolated vortex embedded in an infinite fermionic superfluid, for the same couplings and temperatures considered in Fig.÷\ref{Figure-2}. In each case, the results of solution of the BdG equations obtained in Ref.\cite{SPS-2013} (full lines) are compared with both the LPDA expression (\ref{number-density-general}) (dashed lines) and 
the LDA expression (\ref{coarse-garined-density}) (dotted lines), which includes and neglects the effect of $\nabla \varphi(\mathbf{r})$, respectively.}
\label{Figure-3}
\end{figure}

Figure \ref{Figure-3} shows the radial density profiles (for the same couplings and temperatures considered in Fig.\ref{Figure-2}), obtained both within the LPDA expression (\ref{number-density-general}) and the LDA expression (\ref{coarse-garined-density}), and compares them with those obtained by the full solution of the BdG equations reported in Ref.\cite{SPS-2013}.
One concludes from this comparison that the LPDA approach provides a valuable approximation to the full BdG calculation also as far as the density profiles are concerned, except close to the center of the vortex (say, within $\rho k_{F} \lesssim 1$) on the BCS side of unitarity at low temperatures where deviations from the BdG results appear. 
Note however that, outside this region, the LPDA improves on the comparison with the BdG results with respect to LDA.

The above discrepancies should have been expected from the analysis made in Ref.\cite{SPS-2013}, where it was shown that to obtain accurate values of the density at the center of a vortex one has to account for the detailed structure of the fermionic BdG wave functions belonging to the continuum spectrum \emph{close to threshold}, whose wavelengths are larger than the local variation of the gap parameter.
These are the local fluctuations which cannot be accounted for by the LPDA approach.
However, once the deviation of $n(\rho)$ from its asymptotic value $n_{0}$ is integrated radially up to a maximum value $\rho_{\mathrm{max}}$ (as it is relevant on physical grounds), the above local discrepancies between the LPDA and BdG calculations get considerably reduced, reaching at most $20 \%$ for coupling $(k_{F} a_{F})^{-1} = -1.0$ and zero temperature when $\rho_{\mathrm{max}}$ is of the order of the vortex radius.

% Figure 4
\begin{figure}[t]
\includegraphics[angle=0,width=8.7cm]{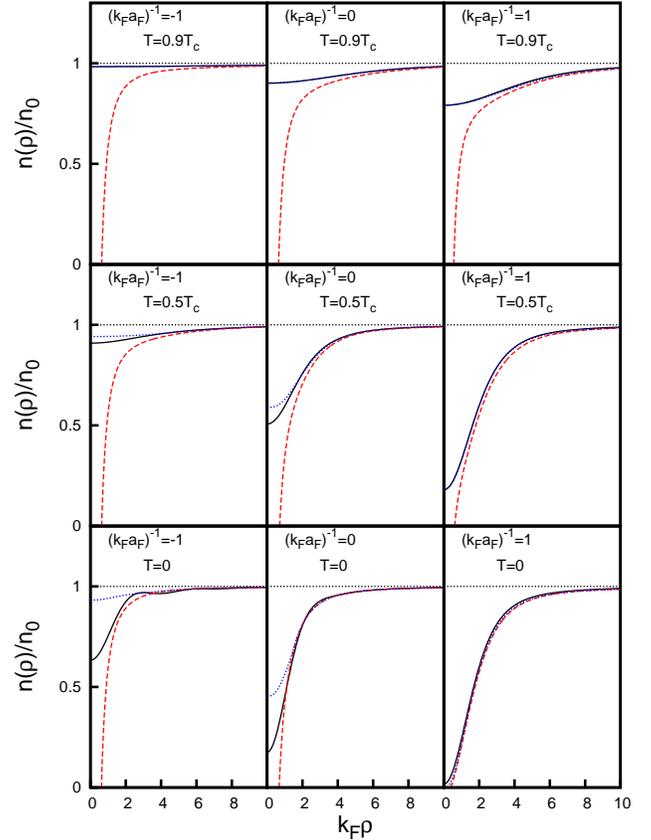}
\caption{(Color online) Radial profiles of the number density for an isolated vortex, for the same couplings and temperatures of Fig.÷\ref{Figure-3}. The results of solution of the BdG equations from Ref.\cite{SPS-2013} (full lines) are compared with the results obtained by the approximate expression (\ref{density-expanded}) (dashed lines) and the LPDA expression (\ref{number-density-general}) (dotted lines).}
\label{Figure-4}
\end{figure}

Outside the center of a vortex, or else in situations where the magnitude of 
$(\nabla \varphi(\mathbf{r})/2 - \mathbf{A}(\mathbf{r}))$ remains small enough (with respect to the inverse of the coherence (healing) length at the given temperature), the right-hand side of Eq.(\ref{number-density-general}) can be expanded to the lowest significant order in $(\nabla \varphi(\mathbf{r})/2 - \mathbf{A}(\mathbf{r}))$, yielding:

\begin{eqnarray}
& & n(\mathbf{r}) \simeq \bar{n}(\mathbf{r}) + \frac{1}{m} \! \left(\frac{\nabla \varphi(\mathbf{r})}{2} - \mathbf{A}(\mathbf{r}) \right)^{2} \! \! \int \! \frac{d\mathbf{k}}{(2 \pi)^{3}}
\nonumber \\
& \times & \left\{\!
\frac{\xi(\mathbf{k}|\mathbf{r})^{2}}{E(\mathbf{k}|\mathbf{r})^{2}} \frac{\partial f_{F}(E(\mathbf{k}|\mathbf{r}))}{\partial E(\mathbf{k}|\mathbf{r})} 
+ \frac{\mathbf{k}^{2}}{3m} \frac{\xi(\mathbf{k}|\mathbf{r})}{E(\mathbf{k}|\mathbf{r})} \frac{\partial^{2} f_{F}(E(\mathbf{k}|\mathbf{r}))}{\partial E(\mathbf{k}|\mathbf{r})^{2}}  \right.
\nonumber \\
& - & \left. \frac{|\Delta(\mathbf{r})|^{2}}{2 \, E(\mathbf{k}|\mathbf{r})^{3}} \left[ 1 - 2 f_{F}(E(\mathbf{k}|\mathbf{r})) \right] \! \right\}
\label{density-expanded}
\end{eqnarray}
\vspace{-0.2cm}

\noindent
where $\bar{n}(\mathbf{r})$ is given by the LDA expression (\ref{coarse-garined-density}) while $\xi(\mathbf{k}|\mathbf{r})$ and $E(\mathbf{k}|\mathbf{r})$ are given by Eq.(\ref{two-definitions}).
Note that when approaching the normal phase whereby $\Delta(\mathbf{r}) \rightarrow 0$, the second term of Eq.(\ref{density-expanded}) vanishes owing to the identity:
\begin{equation}
\int \! \frac{d\mathbf{k}}{(2 \pi)^{3}} \left\{ \frac{\partial f_{F}(\xi(\mathbf{k}|\mathbf{r}))}{\partial \xi(\mathbf{k}|\mathbf{r})} + \frac{\mathbf{k}^{2}}{3m} \,
\frac{\partial^{2} f_{F}(\xi(\mathbf{k}|\mathbf{r}))}{\partial \xi(\mathbf{k}|\mathbf{r})^{2}} \right\} \, = \, 0
\label{first-important-identity}
\end{equation}
\noindent
which holds for a normal system for any value of $\mu(\mathbf{r})$ and temperature.
In this case, $n(\mathbf{r}) \rightarrow \bar{n}(\mathbf{r})|_{\Delta(\mathbf{r})=0}$.

A comparison between the results of the approximate expression (\ref{density-expanded}) and the BdG calculation for the local density is provided in 
Fig.÷\ref{Figure-4}, for the same couplings and temperatures of Fig.÷\ref{Figure-3}.
One sees that the approximate expression (\ref{density-expanded}) is able to reproduce quite well the results of the BdG calculation outside the inner region of the vortex where $|\nabla \varphi(\mathbf{r})|$ remains bounded.

%%%%%%%%%%%%%%  Sub-section 3B  %%%%%%%%%%%%%%%%%%%%%%%% 
\vspace{0.1cm}
\begin{center}
{\bf B. Coarse grained current}
\end{center}
\vspace{0.1cm}

The most general expression that can be written for the current density within the LPDA approach of the main text is as follows:
\begin{eqnarray}
\mathbf{j}(\mathbf{r}) & = & \frac{1}{m} \! \left(\frac{\nabla \varphi(\mathbf{r})}{2} - \mathbf{A}(\mathbf{r}) \right) \, n(\mathbf{r})
\nonumber \\
& + &  2 \int \! \frac{d\mathbf{k}}{(2 \pi)^{3}} \, \frac{\mathbf{k}}{m} \, f_{E} \! \left( E^{\mathbf{A}}_{+}(\mathbf{k}|\mathbf{r}) \right)
\label{number-current-general} 
\end{eqnarray}
\noindent
where the fermion density $n(\mathbf{r})$ is given by Eq.(\ref{number-density-general}) and $E^{\mathbf{A}}_{+}(\mathbf{k}|\mathbf{r})$ by Eq.(\ref{three-definitions}).
At the center of the vortex where $|\Delta(\mathbf{r})| = 0$, the two terms on the right-hand side of Eq.(\ref{number-current-general}) compensate each other making $\mathbf{j}(\mathbf{r})$ vanish.

% Figure 5
\begin{figure}[t]
\includegraphics[angle=0,width=8.7cm]{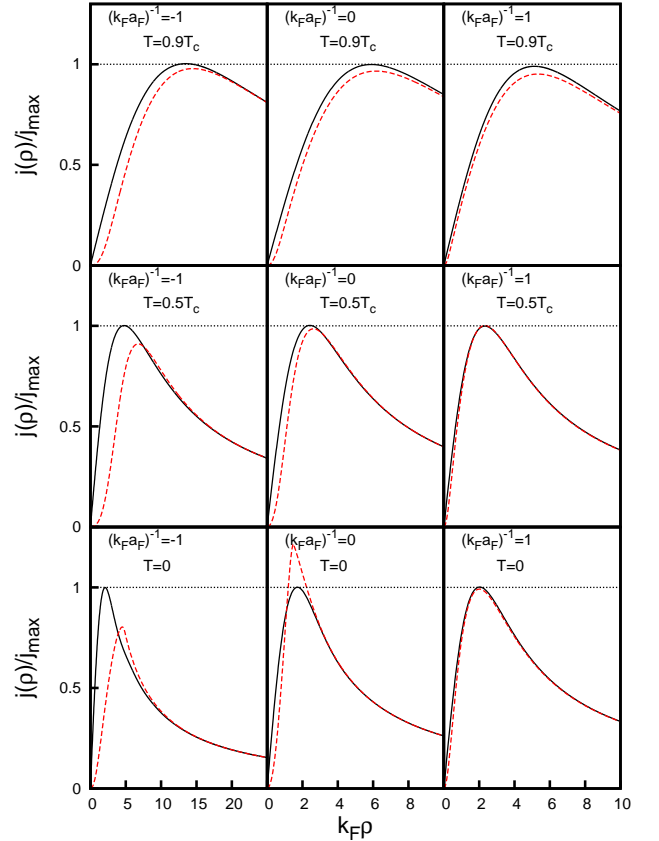}
\caption{(Color online) Radial profiles of the current density $j(\rho)$ for an isolated vortex embedded in an infinite fermionic superfluid, for the same couplings and temperatures of Fig.÷\ref{Figure-2}. The results of the solution of the BdG equations obtained in Ref.\cite{SPS-2013} (full lines) are compared with those obtained by the expression (\ref{number-current-general}) (dashed lines). The maximum value $j_{\mathrm{max}}$ of $j(\rho)$ corresponds to the BdG calculation.}
\label{Figure-5}
\end{figure}

A comparison between the radial profiles of the current density for an isolated vortex embedded in an infinite fermionic superfluid, obtained from the expression (\ref{number-current-general}) and from the BdG calculation of Ref.\cite{SPS-2013}, is shown in Fig.÷\ref{Figure-5} for various couplings and temperatures.
The overall comparison between the two calculations appears quite good, especially regarding the decay of the current past its maximum and also as far as the position of the maximum is concerned.
However, deviations between the LPDA and BdG calculations are more evident for $j(\rho)$ than for $\Delta(\rho)$ in the BCS regime at low temperature and especially near the center of the vortex where the spatial variation $\mathbf{Q}$ of the phase of the order parameter diverges.
Under these circumstances, the LPDA approach tends to suppress locally the superfluid density with respect to the BdG calculation as well as to increase the normal density at the same time (cf. Fig.÷\ref{Figure-3}).

We may also consider an approximate version of the expression (\ref{number-current-general}), which holds when the magnitude of 
$(\nabla \varphi(\mathbf{r})/2 - \mathbf{A}(\mathbf{r}))$ is small enough and is obtained by expanding the right-hand side of Eq.(\ref{number-current-general}) to the lowest significant order in $(\nabla \varphi(\mathbf{r})/2 - \mathbf{A}(\mathbf{r}))$ as follows: 
\begin{equation}
\mathbf{j}(\mathbf{r}) = \frac{1}{m} \! \left(\frac{\nabla \varphi(\mathbf{r})}{2} - \mathbf{A}(\mathbf{r}) \right) \, n_{s}(\mathbf{r})
\label{number-current-expanded}
\end{equation}
\noindent
where 
\begin{equation}
n_{s}(\mathbf{r}) = \bar{n}(\mathbf{r}) + 2 \int \! \frac{d\mathbf{k}}{(2 \pi)^{3}} \frac{\mathbf{k}^{2}}{3 m}
\frac{\partial f_{E}(E(\mathbf{k}|\mathbf{r}))}{\partial E(\mathbf{k}|\mathbf{r})}
\label{superfluid-density}
\end{equation}
\noindent
can be identified as the \emph{local superfluid density} \cite{Sa_de_Melo-2006}.

% Figure 6
\begin{figure}[t]
\includegraphics[angle=0,width=8.7cm]{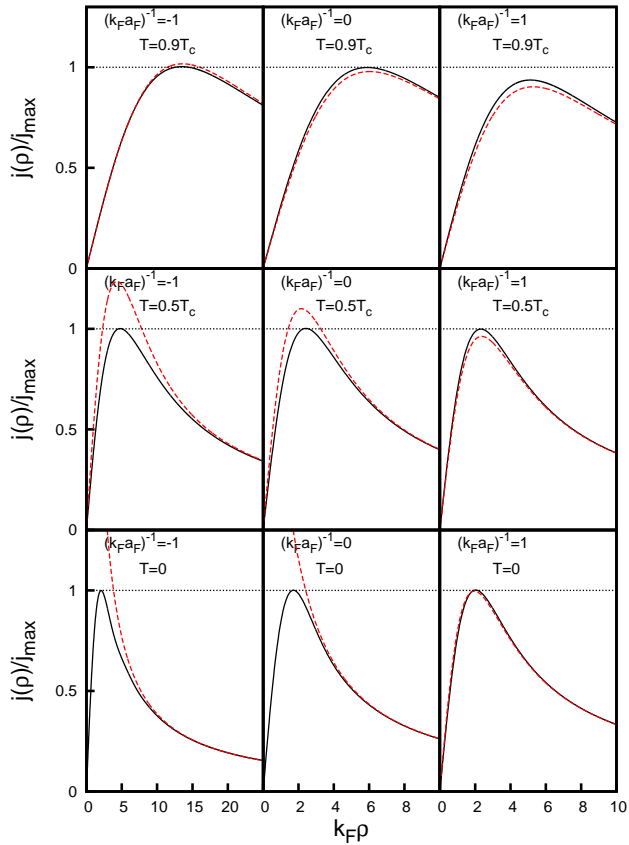}
\caption{(Color online) Radial profiles of the current density $j(\rho)$ for an isolated vortex, for the same couplings and temperatures of Fig.÷\ref{Figure-5}. The results of solution of the BdG equations from Ref.\cite{SPS-2013} (full lines) are compared with those obtained by the approximate expression (\ref{number-current-expanded}) (dashed lines). The maximum value $j_{\mathrm{max}}$ of $j(\rho)$ corresponds to the BdG calculation.}
\label{Figure-6}
\end{figure}

A comparison between the results of the approximate expression (\ref{number-current-expanded}) and the BdG calculation for the current density is provided in Fig.÷\ref{Figure-6}, for the same couplings and temperatures of Fig.÷\ref{Figure-5}.
One again verifies that an approximate expression like (\ref{number-current-expanded}) is able to reproduce the results of the BdG calculation outside the inner region of the vortex where $|\nabla \varphi(\mathbf{r})|$ remains bounded.

Note that when approaching the normal phase whereby $\Delta(\mathbf{r}) \rightarrow 0$, in the second term on the right-hand side of Eq.(\ref{superfluid-density}) one can expand:

\begin{equation}
\frac{\partial f_{E}(E(\mathbf{k}|\mathbf{r}))}{\partial E(\mathbf{k}|\mathbf{r})} \simeq \frac{\partial f_{E}(\xi(\mathbf{k}|\mathbf{r}))}{\partial \xi(\mathbf{k}|\mathbf{r})} +
\frac{|\Delta(\mathbf{r})|^{2}}{2 \xi(\mathbf{k}|\mathbf{r})} \frac{\partial^{2} f_{E}(\xi(\mathbf{k}|\mathbf{r}))}{\partial \xi(\mathbf{k}|\mathbf{r})^{2}}
\label{controlled-expansion}
\end{equation}
\noindent
for which no singularity occurs when $\xi(\mathbf{k}|\mathbf{r}) \rightarrow 0$.
When inserted into Eq.(\ref{superfluid-density}), the first term on the right-hand side of Eq.(\ref{controlled-expansion}) cancels with the first term on the right-hand side of Eq.(\ref{superfluid-density}) with $|\Delta(\mathbf{r})| = 0$ owing to the identity:

\begin{equation}
\int \! \frac{d\mathbf{k}}{(2 \pi)^{3}} \left\{ f_{F}(\xi(\mathbf{k}|\mathbf{r})) + \frac{\mathbf{k}^{2}}{3m} \,
\frac{\partial f_{F}(\xi(\mathbf{k}|\mathbf{r}))}{\partial \xi(\mathbf{k}|\mathbf{r})} \right\} \, = \, 0 
\label{second-important-identity}
\end{equation}

\noindent
which holds similarly to Eq.(\ref{first-important-identity}).
As a result, $n_{s}(\mathbf{r})$ is proportional to $|\Delta(\mathbf{r})|^{2}$, albeit with a coefficient that diverges in the zero-temperature limit 
when $\mu > 0$, as is evident from Fig.÷\ref{Figure-6}.

In the BCS limit close to the critical temperature $T_{c}$ (corresponding to the Ginzburg-Landau regime) only the second term on the right-hand side of Eq.(\ref{controlled-expansion}) contributes to $n_{s}(\mathbf{r})$ by parity arguments about the Fermi surface, yielding:
\begin{eqnarray}
n_{s}^{\mathrm{GL}}(\mathbf{r}) & \simeq & \frac{|\Delta(\mathbf{r})|^{2}}{(2 k_{B} T_{c})^{3}} \int \! \frac{d\mathbf{k}}{(2 \pi)^{3}} \frac{\mathbf{k}^{2}}{3 m} \, 
\left. \frac{\tanh x}{x \cosh^{2} x} \right|_{x=\frac{ \xi(\mathbf{k}|\mathbf{r})}{2 k_{B} T_{c}}}
\nonumber \\
& \simeq & 2 \, |\Delta(\mathbf{r})|^{2} \frac{7 \, \zeta(3) \, n}{8 (\pi k_{B} T_{c})^{2}} = 2 \, |\Psi(\mathbf{r})|^{2} \, .
\label{BCS-superfluid-density}
\end{eqnarray}
\noindent
Here, $n$ is the value of the homogeneous density when $V(\mathbf{r}) = 0$, $\zeta(3)$ is the Riemann zeta function of argument $3$, and 
$\Psi(\mathbf{r}) = \Delta(\mathbf{r}) \sqrt{7 \, \zeta(3) \, n / 8 (\pi k_{B} T_{c})^{2}}$ is the wave function of Cooper pairs in this limit \cite{FW}.

In the opposite BEC limit at low temperature (corresponding to the Gross-Pitaevskii regime), on the other hand, only the first term on the right-hand side of Eq.(\ref{superfluid-density}) contributes to $n_{s}(\mathbf{r})$, and an expansion of the expression (\ref{coarse-garined-density}) to the lowest order in $|\Delta(\mathbf{r})|^{2}$ yields:
\begin{eqnarray}
n_{s}^{\mathrm{GP}}(\mathbf{r}) & \simeq & \bar{n}(\mathbf{r}) \simeq \frac{|\Delta(\mathbf{r})|^{2}}{2} \int \! \frac{d\mathbf{k}}{(2 \pi)^{3}} 
\frac{1}{\left( \frac{\mathbf{k}^{2}}{2m} + |\mu| \right)^{2}} 
\nonumber \\
& = & 2 \, |\Delta(\mathbf{r})|^{2} \, \frac{m^{2} \, a_{F}}{8 \, \pi} = 2 \, |\Phi(\mathbf{r})|^{2}
\label{BEC-superfluid-density}
\end{eqnarray}
\noindent
where $\Phi(\mathbf{r}) = \Delta(\mathbf{r}) \sqrt{m^{2} a_{F} / 8 \pi}$ is the wave function of composite bosons in this limit \cite{PS-2003}.

%%%%%%%%%%%%%%  SECTION 4  %%%%%%%%%%%%%%s 
% Section 4
\section{IV. Concluding remarks and outlook}

In this paper, we have obtained a nonlinear differential (LPDA) equation for the gap parameter by a coarse-graining procedure of the BdG equations, with the aim of speeding up the computer time and reducing the memory space for solving these equations in an effective way when dealing with problems that involve superconducting/superfluid systems in the presence of nontrivial spatially dependent external fields.
In fact, in spite of their apparent simplicity, accurate solutions of the original BdG equations can be obtained at the price of considerable efforts only for a limited number of relatively simple problems (among which one can mention the Josephson flow across a one-dimensional barrier \cite{SPS-2007-2010} and an isolated vortex embedded in an infinite superfluid \cite{SPS-2013}). 

We have also presented favorable a numerical test of the LPDA equation, whereby the solution of the LPDA equation has effectively replaced that of the BdG equations for the case of an isolated vortex. 
From this test one can expect that the LPDA equation could provide accurate enough solutions also in more complicated physical problems for which a direct application of the BdG equations will be out of reach.

In practice, the importance of the proposed method lies in the fact that the LPDA equation has essentially the structure, on the one hand, of the GP equation (to which it reduces in the BEC limit of the BCS-BEC crossover at low temperature) and, on the other hand, of the GL equation (to which it reduces in the BCS limit of the BCS-BEC crossover close to the critical temperature). 
Both equations have, in fact, a long history of \emph{practical applications}, to problems related to dilute bosons at low temperature for the GP equation \cite{PSt-2003} or to strongly-overlapping Cooper pairs close to the critical temperature for the GL equation \cite{Tinkham-1980}.

Quite generally, finding an efficient way of solving the BdG equations by replacing them with the LPDA equation can be relevant not only for problems in condensed matter or in ultracold gases, but also in nuclear physics (including neutron stars) where the BdG equations are better known as the Hartree-Fock-Bogoliubov equations \cite{RS-2004}.
For instance, in condensed matter the method could be applied to superconducting systems with reduced dimensionality and at the nanoscale level also in the presence of quantum confinement, thus bringing out the sensitivity of the superconducting properties on the specimen geometry \cite{Shanenko-2007} or addressing quantum-size effects in the BCS-BEC crossover \cite{Shanenko-2012}.

About the BCS-BEC crossover in ultracold gases, application of the LPDA equation could prove essential to account for the experimental data on the occurrence of arrays of vortices \cite{Ketterle-2005} or the quenching of the moment of inertia \cite{Grimm-2011}, which are of particular importance since they have revealed unambiguously the presence of a superfluid phase at low enough temperature in an ultracold Fermi gas contained in a rotating trap.
These phenomena can also be of interest to nuclear physics, in particular as far as the inner crust of neutron stars is concerned \cite{Broglia-2008,CH-2008}.
In this context, it is worth mentioning a related work done by Bulgac and co-workers through an extension of the Kohn-Sham approach to superfluid Fermi systems that goes even beyond the BdG equations by including correlation effects over and above mean field \cite{Bulgac-2007-2013}.

Still, about rotating traps, it is worth emphasizing the presence of the vector potential in the arguments of the Fermi functions that enter the coefficients of the LPDA equation.
This feature, which distinguishes the present from other proposals also based generically on the slow spatial variation of the gap parameter \cite{Devreese-2013}, is essential to account for
the pair-breaking effects of rotation on ultracold Fermi gases in the BCS-BEC crossover, as already discussed in Ref.\cite{US-2008} although in the absence of vortices.

Further applications of the LPDA equation can be conceived in the context of the Josephson and related effects when using the original BdG equations would be computationally too demanding, to study, for instance, multiple barriers with resonant levels or Josephson coupling between planar superfluids, which can be of cross interest to condensed-matter and ultracold atoms physics also within the BCS-BEC crossover.
Dealing with these phenomena can be considered at finite temperature as well, to assess, for instance, the general validity of the Landau criterion for superfluidity which can also be addressed experimentally with ultracold atoms \cite{Ketterle-2007}.

In addition, extensions of the LPDA approach to spin-imbalanced systems appears feasible along the lines of Ref.\cite{PS-2006} (which has, however, addressed only the limit of a Bose-Einstein condensate); and possibly even to nonequilibrium situations by relying on the Keldysh approach to superconductivity \cite{Rammer-Smith-1986} in the place of the Gor'kov approach that was utilized in the present paper for equilibrium situations.
On the other hand, inclusion of pairing fluctuations beyond mean field as well as of time-dependent effects would most certainly require quite more intense efforts to be implemented.

A comment on the need for including pairing fluctuations beyond mean field is in order. 
It is known that, in general, a correct description (especially at finite temperature) of the physics of the BCS-BEC crossover would require one to include pairing-fluctuations beyond mean field
\cite{Strinati-2012}, as this is certainly the case for homogeneous systems.
However, fluctuation effects are are also known to be in practice less severe for inhomogeneous systems, which are those for which the BdG equations are ideally suited. 
Accordingly, it is then clear that the LPDA equation should most suitably be used to shine light on this kind of complicated inhomogeneous situations, for which implementing the original BdG equations would remain a formidable task while an even further inclusion of fluctuations might be essentially out of reach.

A final comment should be made on the applicability itself of a \emph{local} (differential) equation (like the LPDA equation) in the weak-coupling (BCS) regime when the temperature is much lower than the critical temperature, such that the GL equation does not apply in principle.
As a matter of fact, the explicit comparison we have shown, between the results of the original BdG equations and its approximate version given by the LPDA equation over a wide range of coupling and temperature, confirms the expectation that a local differential approach is bound to fail in the BCS regime at low temperature, owing to the fact that the size of the Cooper pairs is quite large and comparable with that of the solution itself.

To overcome this problem, while abandoning at the same time the full solution of the original BdG equations due to practical complexities, schemes have been devised over the time to trade all the information and details provided by the BdG equations for a reduction of these complexities, yet still sticking to the weak-coupling (BCS) regime \cite{Eilenberger-1968}.
In this context, future work could improve on the comparison with the BdG results at low temperature in weak coupling, by utilizing the non-local (integral) equation 
(\ref{not-yet-final-LPDA-equation}) for the gap parameter in the place of the local (differential) LPDA equation (\ref{LPDA-differential-equation}).
In turn, this non-local equation could be applied, e.g., to problems related to disorder, thereby extending previous approaches \cite{Kogan-1985} away from the weak-coupling limit.

%%%%%%%%%%%%%%%%%%%%%%%%%%%%%%%%%
\vspace{0.1cm}

\begin{center}
\begin{small}
{\bf ACKNOWLEDGMENTS}
\end{small}
\end{center}
\vspace{-0.1cm}

This work was partially supported by the Italian MIUR under Contract Cofin-2009 ``Quantum gases beyond equilibrium''.
The authors are indebted to P. Pieri for a critical reading of the manuscript and for preparing Figs.1, 7, and 8.
 
\vspace{0.4cm}                                                                                                                                                                                                                                                                                                                                                                                                         
%%%%%%%%%%%%%%  APPENDIX   %%%%%%%%%%%%%% 
% Appendix 
\appendix
\section{APPENDIX: COEFFICIENTS $\mathcal{I}_{0}$ AND $\mathcal{I}_{1}$ IN TERMS OF ELLIPTIC INTEGRALS}

In this Appendix, we provide analytic expressions for the coefficients $\mathcal{I}_{0}$ and $\mathcal{I}_{1}$ of the LPDA equation (\ref{LPDA-differential-equation}) in the limit of zero temperature and with $A(\mathbf{r}) = 0$.
Specifically, we show that under these circumstances $\mathcal{I}_{0}$ and $\mathcal{I}_{1}$ can be expressed in terms of elliptic integrals according to the results of Ref.\cite{MPS-1998}.

From their definitions (\ref{I_0-definition-finite_temperature}) and (\ref{I_1-definition-finite_temperature}), we then write:
\begin{equation}
\mathcal{I}_{0} = \frac{1}{2 \pi^{2}} \int_{0}^{\infty} \!  dk \, k^{2} \left( \frac{1}{2 \, \left[ \left( \frac{\mathbf{k}^{2}}{2m} - \mu \right)^{2} + \Delta^{2} \right]^{1/2}} - \frac{m}{ k^{2}} \right)
\label{I_0-zero_temperature}
\end{equation}
\noindent
and
\begin{equation}
\mathcal{I}_{1} = \frac{1}{8 \pi^{2}} \int_{0}^{\infty} \!  dk \, k^{2} 
\frac{\left( \frac{\mathbf{k}^{2}}{2m} - \mu \right)}{\left[ \left( \frac{\mathbf{k}^{2}}{2m} - \mu \right)^{2} + \Delta^{2} \right]^{3/2}} 
\label{I_0-zero_temperature}
\end{equation}
\noindent
where the $\mathbf{r}$-dependence of $\Delta(=|\Delta|)$ and of $\mu$ has been dropped for convenience.
With the same notation of Ref.\cite{MPS-1998}, we then rewrite:
\begin{eqnarray}
\mathcal{I}_{0} & = & \frac{(2m)^{3/2} \, \sqrt{\Delta}}{2 \pi^{2}} \,\,\, \left[ x_{0} \, I_{6}(x_{0}) \, - \,  I_{5}(x_{0}) \right] 
\nonumber \\
\mathcal{I}_{1} & = & \frac{(2m)^{3/2}}{8 \pi^{2} \, \sqrt{\Delta}} \,\,\, I_{6}(x_{0})
\label{I_0-and-I_1-vs-elliptic_integtals}
\end{eqnarray}
\noindent
where $x_{0} = \mu / \Delta$ and 
\begin{eqnarray}
I_{5}(x_{0})  \!\! & = & \!\! (1 + x_{0}^{2})^{1/4} \, E\left(\frac{\pi}{2},\kappa\right) - \frac{1}{4 \, x_{1}^{2} \, (1 + x_{0}^{2})^{1/4}} \, F\left(\frac{\pi}{2},\kappa\right)
\nonumber \\
I_{6}(x_{0}) \!\! & = & \!\! \frac{1}{2 \, (1 + x_{0}^{2})^{1/4}} \, F\left(\frac{\pi}{2},\kappa\right) \, .
\label{I_5-and-I_6-vs-elliptic_integtals}
\end{eqnarray}
\noindent
In these expressions
$x_{1}^{2} = (\sqrt{1 + x_{0}^{2}} \, + \, x_{0}) / 2$ and $\kappa^{2} = x_{1}^{2} / \sqrt{1 + x_{0}^{2}}$, while $E\left(\frac{\pi}{2},\kappa\right)$ and $ F\left(\frac{\pi}{2},\kappa\right)$ are the complete elliptic integrals.

% Figure 7
\begin{figure}[h]
\includegraphics[angle=0,width=8.8cm]{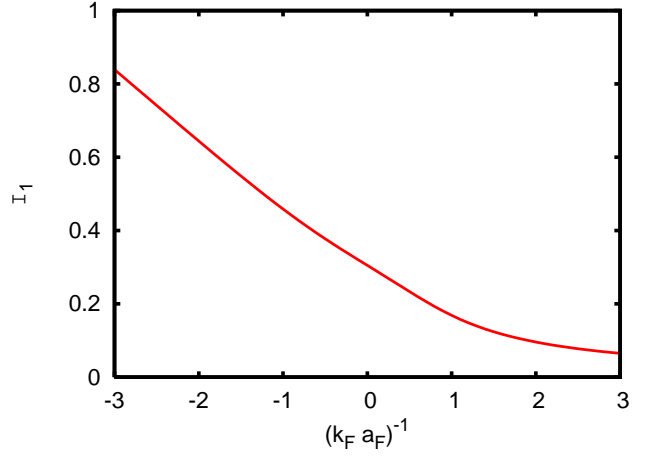}
\caption{(Color online) Coefficient $\mathcal{I}_{1}$ (in units of $2 m^{2}/(\pi^{2} k_{F})$) vs the coupling parameter $(k_{F} a_{F})^{-1}$, obtained from the expression (\ref{I_0-and-I_1-vs-elliptic_integtals}) where the values of $\Delta$ and $\mu$ are taken from mean field at $T=0$.}
\label{Figure-7}
\end{figure}

For the sake of example, a plot of $\mathcal{I}_{1}$ according to the expression (\ref{I_0-and-I_1-vs-elliptic_integtals}) is given in Fig.\ref{Figure-7} vs the coupling parameter $(k_{F} a_{F})^{-1}$, where the values of $\Delta$ and $\mu$ are taken from mean field at $T=0$ (in this case, owing to the BCS gap equation 
$\mathcal{I}_{0}$ equals $-m/(4 \pi a_{F})$ for all couplings).

We are interested, in particular, in what happens near the center of a vortex, whereby $\Delta \rightarrow 0$ and either one of the two limits
$x_{0} \rightarrow + \infty$ and $x_{0} \rightarrow - \infty$ is correspondingly relevant.

In the limit $x_{0} \rightarrow + \infty$, one obtains $I_{5}(x_{0}) \simeq \sqrt{x_{0}}$ and $I_{6}(x_{0}) \simeq \ln(8 x_{0}) / (2 \sqrt{x_{0}})$ \cite{MPS-1998}, from which:
\begin{eqnarray}
\mathcal{I}_{0} & \simeq & \frac{(2m)^{3/2} \, \sqrt{\mu}}{4 \pi^{2}} \,\, \ln \left(\frac{8 \, \mu}{\Delta}\right)
\nonumber \\
\mathcal{I}_{1} & \simeq & \frac{1}{4 \mu} \, \mathcal{I}_{0} \, .
\label{I_0-and-I_1-BCS_limit}
\end{eqnarray}
\noindent
In this limit, the LPDA equation, namely,
\begin{equation}
\left( \frac{m}{4 \pi a_{F}} + \mathcal{I}_{0} \right) \Delta(\mathbf{r}) \, + \, \mathcal{I}_{1} \, \frac{\nabla^{2}}{4m} \Delta(\mathbf{r}) \, = \, 0
\label{LPDA_equation}
\end{equation} 
\noindent
for given value of $a_{F}$ reduces to:
\begin{equation}
4 \, \mu \, \Delta(\mathbf{r}) \, + \, \frac{\nabla^{2}}{4m} \Delta(\mathbf{r}) \, = \, 0
\label{LPDA_equation-simplified-BCS}
\end{equation}
\noindent
so that in this case the relevant length scale for $\Delta(\mathbf{r})$ is the inverse of $k_{\mu} = \sqrt{2 m \mu}$ ($\mu > 0$).
This conclusion was also reached in Ref.\cite{SRH-2006}, while studying the profile of an isolated  vortex at zero temperature directly in terms of the BdG equations.

% Figure 8
\begin{figure}[t]
\includegraphics[angle=0,width=8.8cm]{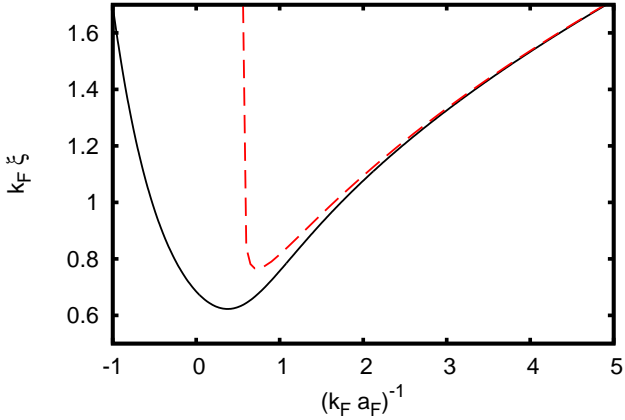}
\caption{(Color online) The coupling dependence of the healing length $\xi$ obtained from the expression (\ref{healing-length-definition}) (and divided by a factor $\sqrt{2}$) (dashed line) is compared with that of the phase coherence length $\xi_{\mathrm{phase}}$ at $T=0$ from Ref.\cite{PS-1996} (full line).}
\label{Figure-8}
\end{figure}

In the opposite limit $x_{0} \rightarrow - \infty$, one obtains instead $I_{5}(x_{0}) \simeq \pi / (16 |x_{0}|^{3/2})$ and $I_{6}(x_{0}) \simeq \pi / (4 |x_{0}|^{1/2})$ \cite{MPS-1998}, from which:
\begin{eqnarray}
\mathcal{I}_{0} & \simeq & - \frac{(2m)^{3/2}}{8 \pi} \,\, \sqrt{|\mu|}
\nonumber \\
\mathcal{I}_{1} & \simeq &  \frac{(2m)^{3/2}}{8 \pi} \,\, \frac{1}{ 4 \, \sqrt{|\mu|}} \, .
\label{I_0-and-I_1-BEC_limit}
\end{eqnarray}

\noindent
In this limit, the LPDA equation (\ref{LPDA_equation}) reduces to:
\begin{equation}
\left( \frac{m}{4 \pi a_{F}} - \frac{(2m)^{3/2}  \sqrt{|\mu|}}{8 \pi} \right) \Delta(\mathbf{r}) + \frac{(2m)^{3/2}}{32 \pi  \sqrt{|\mu|}} \frac{\nabla^{2}}{4m} \Delta(\mathbf{r}) = 0
\label{LPDA_equation-simplified-BCS}
\end{equation}
\noindent
so that in this case the relevant length scale for $\Delta(\mathbf{r})$ can be identified with the healing length $\xi$ given by:

\begin{equation}
\xi^{2} = \frac{ \frac{(2m)^{3/2}}{32 \pi  \sqrt{|\mu|}} \frac{1}{4m} }{ \frac{m}{4 \pi a_{F}} - \frac{(2m)^{3/2}  \sqrt{|\mu|}}{8 \pi} } \, .
\label{healing-length-definition}
\end{equation}

\noindent
In particular, in the BEC limit, whereby $2 \mu = - (m a_{F}^{2})^{-1} + \mu_{B}$ where $\mu_{B}$ is the chemical potential of the composite bosons that form in that limit, the expression (\ref{healing-length-definition}) reduces to $\xi^{2} = (2 m_{B} \mu_{B})^{-1}$ where $m_{B} = 2m$.

Figure \ref{Figure-8} compares, for couplings on the BEC side of unitarity, the values of the healing length $\xi$ obtained from the expression (\ref{healing-length-definition}) with those of the phase coherence length $\xi_{\mathrm{phase}}$ that were obtained in Ref.\cite{PS-1996} at zero temperature by a completely different method.
In this respect, it is rather remarkable to verify how the relatively simple expression (\ref{healing-length-definition}) is able to reproduce $\xi_{\mathrm{phase}}$ essentially down to the coupling $(k_{F} a_{F})^{-1} \approx +1.0$.

%%%%%%%%%%%%%%%%%%%%%%%%%%%%%%%%%
% Bibliography

%%%%%%%%%%%%%%%%%%%%%%%%%%%%%%%%%


\begin{thebibliography}{99}
\bibitem{footnote-negligible} On the right-hand side of Eq.(\ref{LPDA-differential-equation}) a term containing 
                                              $\sum_{i,j=1}^{3} A_{i}(\mathbf{r}) A_{j}(\mathbf{r}) \frac{\partial^{2}}{\partial x_{i} \partial x_{j}} \Delta(\mathbf{r})$ 
                                              has been omitted, under the assumption that $\mathbf{A}(\mathbf{r})^{2}/m$ is negligible in comparison with all
                                              the other relevant energy scales of the problem.
\end{thebibliography}
\end{document}